\documentstyle[11pt,aaspp]{article}

\def\etal{{\it et~al.\ }}
\def\eg{{\it e.g.\ }}
\def\ie{{\it i.e.\ }}

\def\dn+{\delta n_+}

\def\minmag{\lower.5ex\hbox{$\; \buildrel < \over > \;$}}
\def\magmin{\lower.5ex\hbox{$\; \buildrel > \over < \;$}}
\def\gtwid{\mathrel{\raise.3ex\hbox{$>$\kern-.75em\lower1ex\hbox{$\sim$}}}}
\def\ltwid{\mathrel{\raise.3ex\hbox{$<$\kern-.75em\lower1ex\hbox{$\sim$}}}}
\def\ref{\par\noindent\hangindent.5in\hangafter=1}
\newcommand{\beq}{\begin{equation}}
\newcommand{\eneq}{\end{equation}}
\newcommand{\beqar}{\begin{eqnarray}}
\newcommand{\eneqar}{\end{eqnarray}}
\newcommand{\barn}{\begin{eqnarray*}}
\newcommand{\earn}{\end{eqnarray*}}

\lefthead {Pietrini and Krolik}
\righthead {}

\begin{document}
\footnotesize
\title{The Inverse Compton Thermostat in Hot Plasmas Near Accreting Black
Holes}

\author{Paola Pietrini}
\affil{ Dipartimento di Astronomia e Scienza dello Spazio, Universit\'a
        di Firenze, Largo E. Fermi 5, 50125 Firenze, Italy}
\and
\author{Julian H. Krolik}

\affil{Department of Physics and Astronomy, Johns Hopkins University,
	   Baltimore, MD 21218}

\begin{abstract}

    The X-ray spectra of accreting black hole systems generally
contain components (sometimes dominating the total emission) which are well-fit
by thermal Comptonization models with temperatures $\sim 100$ keV.  We
demonstrate why, over many orders of magnitude in heating rate and seed photon
supply, hot plasmas radiate primarily by inverse Compton scattering,
and find equilibrium temperatures within a factor of a few of 100 keV.  We
also determine quantitatively the (wide) bounds on heating rate and seed photon
supply for which this statement is true.

    Plasmas in thermal balance in this regime obey two simple scaling
laws: $\Theta \tau_T \simeq 0.1 (l_h/l_s)^{1/4}$; and $\alpha \simeq 1.6
(l_s/l_h)^{1/4}$.  Here the hot plasma heating rate compactness is $l_h$,
the seed photon compactness is $l_s$, the temperature in electron rest
mass units is $\Theta$, and the Thomson optical depth is $\tau_T$.  The
coefficient in the first expression is weakly-dependent on plasma
geometry; the second expression is independent of geometry.  Only when
$l_s/l_h$ is a few tenths or greater is there a weak secondary dependence
in both relations on $\tau_T$.

   Because $\alpha$ is almost independent of everything but $l_s/l_h$,
the observed power law index may be used to estimate $l_s/l_h$.  In both
AGN and stellar black holes, the mean value estimated this way is
$l_s/l_h \sim 0.1$, although
there is much greater sample dispersion among stellar black holes than
among AGN.  This inference favors models in which the intrinsic (as opposed
to reprocessed) luminosity in soft photons entering the hot plasma
is small, or in which the hard X-ray production is comparatively distant
from the source of soft photons.  In addition, it predicts that $\Theta
\tau_T \simeq 0.1$ -- 0.2, depending primarily on plasma geometry.
It is possible to construct coronal models
(\ie models in which $l_s/l_h \simeq 0.5$) which fit the observed spectra, but
they are tightly constrained: $\tau_T$ must be $\simeq 0.08$ and $\Theta \simeq
0.8$.

\end{abstract}

\section{Introduction}

    Ever since the paper by Shapiro \etal (1976), it has been a commonplace
to explain the hard X-ray spectra of accreting black hole systems,
whether they are galactic binaries or AGN, by thermal
Comptonization.  Indeed, such models often give very good descriptions of
the broad band X-ray spectra of these systems (Sunyaev \& Titarchuk 1980;
Grebenev \etal 1994; Johnson \etal 1994; Zdziarski \etal 1994;
Haardt \etal 1993).
In recent years ART/P, SIGMA and OSSE observations
have substantially increased the volume of data available for constraining
these models, but the basic framework has been sustained: for all systems
identified on the basis of independent arguments as accreting black holes,
thermal Comptonization is an acceptable description of the X-ray
spectrum from $\simeq 3$ -- 300 keV to within the uncertainties in both
measurements and theory.

   The best fit values for the model's two free parameters, temperature
and optical depth, are almost always in the range 30 -- 300 keV and
0.1 -- 5, respectively (Haardt \etal 1993; Bouchet
\etal 1993; Johnson \etal 1993; Zdziarski \etal 1994).  There is certainly no
selection effect
that could explain this comparatively narrow range of parameters in the
AGN, for most are discovered by optical, ultraviolet, or radio searches.
While galactic black hole candidates are almost all discovered by X-ray
techniques, these searches would have been sensitive to a much wider range of
temperatures: for fixed luminosity and optical depth, many of these
sources could
easily have been detected if their temperature had been as low as $\sim 1$ keV
or as high as $\sim 1$ MeV.

   In most previous work in this field, X-ray spectra fit with thermal
Comptonization models
are described in terms of the inferred temperature and optical depth.
For many purposes this is a reasonable procedure, for the spectral
formation models are most directly framed in terms of these parameters,
and they do have a direct physical meaning.  However, if we are to
understand {\it why} these conditions are found, we will do better
to speak in terms of heating rates, cooling rates, and dynamics.  In
lower temperature plasmas, studies of how dissipative heating
and photoionization
balance recombination, bremsstrahlung, and collisional excitation led to
the realization that the dissipation rate and ionization parameter
determine the temperature and ionization state.  In high temperature
plasmas, dissipation and Compton recoil or pair production using
externally-created high energy photons balance annihilation, bremsstrahlung,
and
inverse Compton scattering of low energy photons; it is the primary goal of
this
paper to learn how this balance scales with the externally-imposed parameters
and boundary conditions.

   A principal result of this program will be to explain why it is that
inverse Comptonization dominates other radiation processes in these systems,
and why its parameters are restricted to such a comparatively narrow
range.  We will argue that if one frames the issue in terms of the
physically causal parameters, namely the heating rate and the seed
photon supply rate, the
observed parameters are approximate ``fixed points" of the nonlinear
equations which determine them.   This approach also turns out to have
a second important consequence: when attention is restricted to values of
temperature and optical depth which are actual equilibria, simple
scaling laws emerge relating observables to the causal parameters.

We begin by
outlining the physical processes which operate in these systems (\S 2), and
then describe how we compute equilibria (\S 3).  In \S 4 we show
quantitatively over how wide a range of causal parameters inverse
Comptonization
dominates the cooling rate.  We also show that the equilibrium
temperature does not depend in any direct way on the traditional Compton
$y$-parameter, but that a different scaling relation does describe
the equilibrium quite well. In addition we discover a simple relation
between the power-law index and the ratio of soft compactness to hard.  We then
discuss in \S 5 observable consequences of
the scaling relations, and why the causal parameters for accreting
black holes almost always lie within the (large) volume of parameter
space corresponding to the narrow range of observed parameters.
We summarize in \S 6.

\section {Thermodynamics of Hot Plasmas}

  To highlight the basic physics of these hot plasmas, we focus on a highly
idealized model (following, \eg Bj\"ornsson \& Svensson 1991, BS91 hereafter):
a spherical, homogeneous cloud of pure ionized hydrogen in which all
particle distribution functions are (relativistic) Maxwellians at the
same temperature $\Theta\equiv kT/m_e c^2$ (see Pietrini \& Krolik 1994, PK94
hereafter).
This temperature is maintained by a heating rate per unit volume $\dot H =
3 L_h/(4\pi R^3)$, where $R$ is the radius of the cloud and $L_h$ is its
luminosity. Photons with initial energy $x_0 m_e c^2$ ($x_0 \ll \Theta$)
are supplied externally in such a way that the photon intensity---at $x_0$
and all other photon energies $x m_e c^2$---is isotropic and constant
at every point in the cloud.

Two sorts of equilibria must be established: between
pair production and annihilation; and between heating and cooling.  Because
the cooling rate depends on the number density of leptons,
while the rate of pair
production depends on the number density of high energy photons,
these two equilibria
are strongly coupled.  Although (approximate) fully self-consistent solutions
have been obtained for the version of this problem without an independent
source of soft photons (BS91), previous treatments of equilibria in
which
externally generated soft photons enter the Comptonizing cloud
(Zdziarski 1985; Haardt \& Maraschi 1991, 1993; Ghisellini
\& Haardt 1994) have a number of limitations, some of which we remove in the
present work.  For example, Haardt \& Maraschi
(1991, 1993) have modeled
the production of X-rays in AGNs with a system in which a hot corona of
total (\ie electron plus positron) Thomson optical depth $\tau_T < 1$ covers
an accretion disk.  In their model the entire accretion power of the disk is
dissipated in the corona.  Because the disk thermalizes and reradiates the
majority of the X-rays directed toward it, the ratio between the seed
photon luminosity $L_s$ and the corona luminosity $L_h$ is fixed at
approximately 0.5.
We extend the range of possibilities to include both larger optical depths
and other ratios between the seed photon luminosity and the plasma luminosity.
In addition, their thermal balance does not include annihilation losses,
internal photon production in the corona, or the Comptonization losses on
the internally-generated photons. In their limited range of parameters, these
approximations are justified, but we wish to study these physical
circumstances over a wider range of parameters.
On the other hand, Ghisellini \& Haardt (1994) cover a
much larger range in the luminosity ratio $L_s/L_h$, but they assume a
pure pair plasma, and so are limited to follow a narrow track in heating
rate and optical depth.

\subsection{Key parameters}

   Lightman (1982) and Svensson (1982) showed that it is a very good
approximation to suppose that pair balance equilibria depend on the net
lepton optical depth $\tau_p$ rather than on the net lepton density or
size of the plasma cloud separately.  Note that in the pure hydrogen
limit, $\tau_p = n_p \sigma_T R$, where $n_p$ is the proton density.

   Because Svensson (1984: S84) also showed that the number of
pairs $z = n_+/n_p$, where $n_+$ is the positron number density, could be found
as a function
of $\Theta$ and $\tau_p$, most treatments since have taken $\Theta$ as
one of the independent variables of the system.  In fact, for a given fixed
value of $\Theta$, the pair balance equation, $\dot n_+ = \dot n_A$ (where
$\dot
n_+$ and $\dot n_A$ are the total pair production rate and the pair
annihilation rates, respectively), can be solved to give  $z = z (\Theta,
\tau_p)$, independently of the thermal balance of the system, \ie, without
taking care of how much heating is actually required to maintain this
temperature.  Moreover, with this approach, one finds that there are either
two distinct equilibrium solutions or no solutions at all
for a given choice of  $(\Theta, \tau_p)$, thus implying the existence of
two equilibrium ``branches'' for the equilibrium curve $z= z(\Theta)$ for
a given $\tau_p$ value (see S84, BS91, PK94).
One branch corresponds to solutions for which $z \gtwid 1$, and is
determined essentially by photon (pair-creation) processes; the other
branch is controlled (mostly) by particle processes, and $z \ll 1$.
In a $z=z(\Theta)$ plot the two branches connect at a maximum allowed value for
the plasma temperature, $\Theta_{max} (\tau_p)$; at greater
temperatures the pair production rate always exceeds the annihilation
rate, so that no equilibrium can be found.

Once the pair balance has
been found, for a given $\Theta$, the heating rate $\dot H$ (per unit volume)
required to maintain
that temperature can be determined by equating it to the sum of all
the cooling rates, which can be written as functions of $\Theta$, $\tau_p$,
and $z$.
    In causal terms, however, the temperature is a derived parameter
which depends on $\dot H$, therefore we can choose to take $\dot H$
as the parameter and derive consequently the temperature, together with
$z$, by coupling the thermal balance equation to the pair balance one.
As  pointed out in BS91,
the thermal balance equation
can be written most concisely in terms of the compactness
\beq
l_h = {\sigma_T\over m_e c^3}{4\pi R^2\over 3}\dot H.
\eneq
Thus, for a given value of $\tau_p$, this equation and the pair balance
equation
may be inverted to find $\Theta (l_h,\tau_p)$ and $z(l_h,\tau_p)$.  These
solutions are single-valued, and, therefore more useful than the ones in
terms of $\Theta$.

  Our work differs from that of Svensson in adding a new element,
external soft photon input.  Two more parameters (or more) are required
to specify this.  We suppose, for simplicity, that the soft photon spectrum
is a $\delta$-function at energy $x_0$.  We call their (constant) number
density $n_0$; the corresponding dimensionless quantity is the ratio
$n_0/n_p$.  It is useful to also treat the density of soft photons in
terms of the alternative dimensionless quantity, the soft photon compactness,
\beq
l_s = {\sigma_T\over m_e c^3}{L_s\over R},
\eneq
where $L_s$ is the luminosity of soft photons necessary to maintain the
density $n_0$.  The actual value of $L_s$ depends on geometry; we assume
that it is the luminosity which exactly balances the rate at which soft
photons leave the cloud:
\beq
L_s = 4\pi R^2 \int_{-1}^0 \, d\mu \, \mu I_s = 4 \pi^2 R^2 I_s,
\eneq
where the intensity of soft photons is given by
\beq
I_s (x) = {c \over 4\pi} m_e c^2 x n_0 \delta (x - x_0).
\eneq
Thus, the soft photon compactness definition (eq.~(2)) reduces to
\beq
l_s = \pi \tau_p x_0 {n_0 \over n_p}.
\eneq

In summary, the equilibrium conditions of our system are determined by
four key causal parameters:
\begin{description}
\item  1)~~cloud (hard) compactness ~~~~$l_h$;
\item  2)~~cloud net lepton  (or, equivalently, proton)
		   Thomson optical depth~~~~ $\tau_p$;
\item  3)~~external soft photon compactness ~~~~$l_s$ (or $n_0/n_p$);
\item  4)~~soft photon energy ~~~~$x_0$.
\end{description}

For a given choice of these, from the solution of the coupled pair and thermal
balance equations one derives the other key parameters:
\begin{description}
\item a)~~plasma temperature ~~~$\Theta= \Theta( l_h, \tau_p,l_s, x_0)$,
\item b)~~pair density, $z \equiv n_+/n_p~~\Rightarrow~~$ total optical
depth to scattering $\tau_T\equiv \sigma_T (2n_++n_p)R= (2z+1)\tau_p$;
\end{description}

We stress that over the very large volume of parameter space we
consider solutions parameterized by $l_s/l_h$ do not depend strongly on $x_0$.
This is because Comptonization is unsaturated (in the sense that over
the great majority of the parameter space of interest, most of
the injected photons are not scattered all the way up to the electron
temperature), so that the Compton
cooling rate depends more directly on the energy density of seed photons
than on their number density. Therefore, we can focus our attention
on the first three  key parameters mentioned above and their
interrelations.

\subsection{The Compton Parameter $y$}

   It has long been conventional to describe the results of Comptonization
by a single parameter, which we call
\beq
y_{nr} = 4\Theta \max (\tau_T,\tau_T^2),
\eneq
where $\max(\tau_T,\tau_T^2)$ is an estimate of the mean number of scatterings
a photon suffers before escape, and $4\Theta$ is the
the mean amplification factor, that is the average fractional
photon energy increase  due to a scattering event with an electron drawn
from a Maxwellian distribution at non-relativistic
temperature $\Theta$ (when the electron energy is much larger than that of
the photon).
  This $y_{nr}$ parameter  (the subscript {\it nr} means non relativistic)
arises as the key
dimensionless quantity when one writes a Fokker-Planck equation for the
evolution of photon energies as a result of Comptonization (Kompaneets 1957;
Rybicki \& Lightman 1979).

  In certain commonly encountered limits, it has several important physical
interpretations.  In the limit in which $\Theta \ll 1$ and $\tau_T \gg 1$,
the shape of the emergent spectrum may be written in terms
of $y_{nr}$ (Illarionov \& Sunyaev 1975, Sunyaev \& Titarchuk 1985).
When $\Theta \ll 1$ and $y_{nr} \ll \ln (3\Theta/x_0)$, the average ratio
between the photons' energy on escape and their energy upon entry is simply
$\exp(y_{nr})$ (Illarionov \& Sunyaev 1975;
Shapiro \etal 1976).  Sometimes (\eg Liang 1979) this is taken to
mean that the ratio of the emergent luminosity to the seed luminosity is
$\exp(y_{nr})$, but this is not correct.
As pointed out by Sunyaev \& Titarchuk (1985) and
Dermer \etal (1991), the output luminosity is usually dominated by
photons on the tail of the energy change distribution, not the photons
whose energy change is near the mean (see also \S 4.4.2).
It also takes more time than
average for these higher energy photons to emerge because they must
scatter more times to reach those energies.  In addition, when the
output spectrum is ``hard" enough, the photon energy range which dominates
the luminosity approaches the energy $\Theta$, so that saturation effects
can be important for them, even if they are not important for the average
photon.  For all these reasons, $\exp(y_{nr})$ is not a very good
expression for the relation between $L_h$ and $L_s$.  In \S 4.4.3 we
derive a much more accurate scaling law.

   There are other difficulties which also limit the utility of the
$y$-parameter for the problem we consider.  When
$\Theta$ is not small compared to unity, photon energies do not change
by small fractional amounts per scatter, and a Fokker-Planck description
is therefore of questionable validity.
As for the relation of $y$ with the spectral shape, this turns out
to be quite problematic as well; in fact,
even when $\tau_T \gtwid 3$, we often deal with transrelativistic
or mildly relativistic temperatures, bringing back the
issue just discussed above. On the other hand,
we are also interested in cases in which $\tau_T$ is not very large,
and for values of $\tau_T\sim 1$ or smaller,
the diffusion approximation breaks down, and the spectral
shape is not directly related to $y$ (see \S~3.1.1), even in
principle.

  Despite these difficulties, several attempts have been made to find
generalizations of the $y$-parameter to relativistic temperatures.
All the different attempts agree about how to generalize the expression
for the mean photon amplification per scattering to relativistic temperatures:
\beq
A_r = 1 + 4\Theta {K_3(1/\Theta)\over K_2(1/\Theta)} =
     1 +  4\Theta {K_1 (1/\Theta) \over K_2 (1/\Theta)} + 16 \Theta^2,
\eneq
where $K_n$ is the $n$th-order modified Bessel function, and,
following S84, its usual approximation is
\beq
A_r \simeq 1 + 4\Theta +16\Theta^2,
\eneq
at least for not too large values of $\Theta$.

However, there is significant disagreement about the best way to generalize
the $y$ parameter. Loeb, McKee, \& Lahav (1991) make
the hypothesis that the probability distribution for the number of
scatterings that a photon can suffer is a Poisson distribution.
They then find that the ratio of the average energy of the escaping
photon to its initial value is, similarly to the non-relativistic
case, $\exp(y_{tot})$, where the parameter $y_{tot} = (A_r - 1) <u>$,
and $<u>$ is the average number of
scatterings for a photon before escape.
Therefore, in this view, the representative relativistic Compton parameter can
be
defined as $y_r\equiv y_{tot}$, so that, approximating $A_r$ with
eq.~(8),
\beq
y_r\simeq (4\Theta +16\Theta^2)\tau_T \left (1+{\tau_T\over 3}\right),
\eneq
where $ <u>\simeq \tau_T(1+\tau_T/3)$ is also a reasonable approximation.

On the other hand, Zdziarski \etal (1990) estimate the ratio of escape energy
to initial energy is by supposing that every photon escapes after suffering the
mean number of scatters.  With this approximation, the relativistic
Compton parameter is
\beq
y_* = \ln(1 + 4\Theta + 16\Theta^2) \tau_T (1 + \tau_T/3).
\eneq
where the degree of approximation for $A_r$ is the same as in eq.~(9).

In the range of physical conditions for the Comptonizing plasma
we are interested in,  the two different generalizations discussed
above give rather different estimates for the parameter, and, in
practice we find that neither of them really directly helps describing
even qualitative features of Thermal Comptonization  in the
mildly relativistic temperature regime, as we discuss in \S~4.4.2.

As a matter of fact, in this regime, for even more reasons than
those discussed above for the general case, the use of a Compton parameter
for a description of the conditions of thermal Comptonization
is not much justified. This is probably true for what regards the
characterization of the spectral shape as well, for which the relevance
of the Compton parameter is actually still an open question (see
Titarchuk 1994, and Zdziarski, private communication).

\section{Calculation}

As we remarked before, we must solve two equilibrium conditions, one for
pair balance and one for thermal balance.  Here we describe the physical
processes we included, the approximations we made, and how we arrived
at our solutions.

\subsection{Pair balance}

We write the pair balance equation as follows:
\beq
\dot n_A=(\dot n_+)_{\gamma\gamma} +(\dot n_+)_{\gamma e}+
		 (\dot n_+)_{\gamma p},
\eneq
where the annihilation rate per unit volume $\dot n_A$ is equated to the sum of
the relevant pair production rates, namely
photon-photon (indicated by the subscript $\gamma-\gamma$),
photon-electron ($\gamma-e$), and photon-proton ($\gamma-p$).
Particle-particle pair creation rates are neglected here, since they are always
negligible
(see PK94 and references therein).  We also assume that no pairs
escape from the cloud.

All of the pair creation rates included depend on the  total
Comptonized radiation field in the plasma cloud, which is given by the sum
of the
contribution due to Comptonization of the internally generated photons, and
that coming from Comptonization of the externally created soft
photons. For this reason, we must define a representation for both
contributions.

\subsubsection {Comptonized Radiation Field }

Following S84, we approximate the contribution from
Comptonization of the photons originating in the cloud plasma
with the photon number density spectral distribution
\beq
n_c (x) = n_W(x)+ n_F(x) =\left [ {n_{\gamma}\over 2}{x^2\over \Theta^2}
		{{\rm e}^{-x/\Theta}\over \Theta}\right ] +
		\left [n_1{{\rm e}^{-x/\Theta}\over x}\right ].
\eneq
$n_c(x)$ is the sum of a Wien spectral component, $n_W(x)$, and a ``flat"
component,
$n_F(x)$.  The factor $n_1$ is the flat component normalization,
explicitly defined in BS91;
$n_\gamma$ is the number density of Wien photons from Comptonization of
internally generated radiation,
and it is derived as
\beq
n_{\gamma}=t_{esc}(f_{B}\dot n_{Br}+f_{DC}\dot n_{DC}),
\eneq
where $\dot n_{Br}$ and $\dot n_{DC}$ are the energy integrated
photon production rates for bremsstrahlung and
for Double Compton process respectively, explicitly defined in Svensson
(1984) and in PK94; $f_{B}$ and $f_{DC}$
( again defined in S84)
are the probabilities that a soft photon of bremsstrahlung or
Double Compton origin is scattered into the Wien peak before escaping
the cloud; these also depend on the parameter $y_1$ defined in S84 which
describes the approach to Wien equilibrium for internally-generated
photons.
Therefore, $n_1$ and $n_{\gamma}$ are functions of $\Theta$ and $\tau_T$.
The photon escape time from the cloud, $t_{esc}$, is a useful quantity
to give a rough description of radiative transfer effects, and it is here
defined as in BS91,
\beq
t_{esc}(\tau_T,\Theta)=\left ({R\over c}\right)\left (1+{1\over
3}g_{\tau}\tau_T                                                \right ),
\eneq
where $g_{\tau}(\Theta)\tau_T= \tau_w$ is the scattering optical
depth including the Klein-Nishina cross section decline at high energies,
by averaging its effects on a Wien photon distribution, and
$g_{\tau}(\Theta)$ is (from BS91)
$$
g_{\tau}(\Theta) = \cases{ (1+5\Theta+0.4\Theta^2)^{-1}&
		   for $\Theta \leq 1$\cr
						      (3/16)\Theta^{-2}(\ln(1.12\Theta)+3/4)[1+(0.1/\Theta)
                  ]^{-1}& for $\Theta\geq 1$\cr};
		  $$
		  the factor 1/3 is related to the spherical geometry chosen (see BS91).

There is another radiation field related relevant quantity that must
be determined, and it is the energy $x_m$ below which photons get strongly
absorbed; this, in general,
represents  the lower limit for spectral integrations.
As a matter of fact, this energy is defined through a relation which is coupled
to the pair  and thermal balance (since it depends on $\Theta$ and $z$)
and must be actually solved simultaneously with those conditions (see
PK94).

Now we consider the contribution to the radiation
field inside the cloud due to Comptonization of the
soft photons originally created by an external source.
Since we are analyzing the problem in the framework
of the thermal pair equilibrium model originally studied by
S84 and BS91, we can be
satisfied with a description of the Comptonized external soft photon
contributions to the radiation field which is at the same level
of approximation as the one we use for the Comptonized internally
generated photons.
A representation of this type is given by Zdziarski (1985, herafter Z85;
1986): the Comptonized ``soft photon''  spectrum is given, in
terms of spectral photon number density $n_{sc}(x)$,  as
the superposition of a power law with an exponential cutoff at the plasma
temperature $\Theta$ and a
Wien peak at the same temperature,
\beq
n_{sc}(x) = n_{0p}\left\{ {1\over 2}\left ( {x\over \Theta}\right )^{-\alpha}
			{e^{-x/\Theta}\over x}+ {1\over 2} \left ({n_{0w}\over n_{0p}}
			\right )
             \left ( {x \over \Theta}\right )^3{e^{-x/\Theta}\over x}
			 \right\}= n_{pl}(x) + n_{sw}(x).
\eneq
In this relation $n_{0w}$ represents the Wien photon density from
Comptonization of external soft photons, and $n_{0p}$ is a normalization
factor for the power-law component.  The relation that best defines
the spectral index $\alpha$ of the power-law component in terms of
$\Theta$ and $\tau_T$ depends on $\tau_T$ (Z85, and corrections
in Zdziarski 1986):
\begin{description}
\item {$\bullet$}~~For $\tau_T$ larger than a few, $\alpha$ is well
described by a relativistic generalization (Z85) of the non-relativistic
form given by Sunyaev \& Titarchuk (1980):
$$
\alpha = \left ( {9\over 4} + \gamma (\tau_T, \Theta)\right )^{1/2}-{3\over 2},
$$
where $\gamma(\tau_T, \Theta)$ is given by Z85,
$$
\gamma (\tau_T, \Theta) = {\pi^2\over 3 (\tau_T+2/3)^2\Theta g_{\Theta}},
$$
with the relativistic correction
$$
g_{\Theta} \equiv K_3(1/\Theta)/K_2(1/\Theta),
$$
again given  in Z85.
\item {$\bullet$}~~For small $\tau_T$, the best description is a
generalization by Z85 to general $\Theta$ and $\tau_T \sim 1$ of the
formula originally derived by Pozdnyakov \etal (1977) for cases with
$\tau_T \ll 1$ and $\Theta \gg 1$:
\beq
\alpha = - {\ln P_{\tau_T}\over \ln A},
\eneq
where
\beq
A = 1+4\Theta g_{\Theta},
\eneq
and
\beq
P_{\tau_T} = 1 - {3\over 8\tau_T^3}[ 2\tau_T^2-1 + e^{-2\tau_T}
			   (2\tau_T+1)],
\eneq
(corrected for a misprint, as explained in Z86).
\end{description}
As for the ratio $(n_{0w}/n_{0p})$, this is given by Z85  as
\beq
{n_{0w}\over n_{0p}}= {\Gamma(\alpha)\over \Gamma
(2\alpha+3)}P_{\tau_T}(\tau_T),
\eneq
 which is again a generalization of ST80 results.

We have now to relate the normalization factor $n_{0p}$ to the
total number density of incoming external soft photons, $n_0$.
{}From the conservation of photon number, in the absence of absorption
processes, we have that
\beq
n_0(1-e^{-\tau_T})= \int_{x_0}^{\infty} n_{sc}(x) dx =
				   {n_{0p}\over 2}\left\{ I_1 +\left (n_{0w}\over
				   n_{0p}\right)I_2\right\},
\eneq
where
$$
I_1 \equiv \int_{x_0/\Theta}^{\infty}t^{-(\alpha+1)} e^{-t} dt
$$
and $I_2 \simeq 2$.
The lower limit of integration chosen, $x_0/\Theta$, implies that we neglect
recoil effects in the Comptonization process.
The integral $I_1$ can be calculated in terms of Whittaker functions,
following Gradhsteyn \& Ryzhik (1980) (integral 3.381.6), and then, since
$x_0/\Theta << 1$, analytically approximated, by using expressions
from Abramowitz \& Stegun (1972) (par. 13.1.33, 13.5.5-13.5.12), so that we
finally obtain
$$
I_1\simeq {1\over\alpha}\left ({x_0\over \Theta}\right)^{-\alpha}
		  e^{-x_0/\Theta},
$$
which is valid for any value of $\alpha$ ($<$ or $\geq$ 1), and
\beq
\left ({n_{0p}\over n_p}\right)= \left({n_0\over n_p}\right)
      (1-e^{-\tau_T}){2\over I_1+2(n_{0w}/n_{0p})}.
\eneq

When $\Theta \ll 1$ and $\tau_T \gg 1$ (where the ST80 approximations
are valid), these approximations are fairly good fits to the ST80
solutions, particularly
at the low and high energy extremes.  However, they underestimate
the ST80 photon intensity by factors of a few in the neighborhood of
$x \sim \Theta$.  When $\tau_T \ll 1$, these approximations agree
reasonably well with Monte Carlo simulations (Zdziarski 1986).  Pair production
rates computed on the basis of these approximations should always
be fairly accurate because they are most sensitive to the highest
energy photons.

\subsubsection{ Radiative Pair Production Rates}

Having made these choices for the representation of the contribution to
the radiation field in the cloud due to Comptonized external
soft photons, we now describe how we calculated the corresponding
contributions to
photon-photon and photon-particle pair production rates.

   All pair production rates involving photons make use of the Wien
component, so our general procedure is to substitute $n_\gamma + n_{0w}$
for $n_\gamma$ in the appropriate expressions in Appendex B of S84.
This is a simple substitution for the photon-particle pair production
rates, and for all those contributions to the photon-photon pair
creation rate except the interaction of Wien photons with the power-law
part of the external photon Comptonized spectrum.  For this last part,
from the general integral form given by S84,
we use the expression
\beq
(\dot n_+)_{wpl} ={cr_e^2\pi\over
2}n_{0p}(n_{\gamma}+n_{0w})\Theta^{\alpha-3}\int_1^{\infty}\phi(s_0)
s_0^{-(\alpha+3)/2} K_{\alpha+3}(2s_0^{1/2}/\Theta)ds_0,
\eneq
where $\phi(s_0)$ is a function given by Gould \& Schreder (1967),
with corrections from Brown, Mikaelian, \& Gould (1973).
Thus, the pair creation rate depends on the strength of Comptonization,
so that the pair balance, thermal balance, and emergent spectrum
calculations must be performed self-consistently.

\subsection{Thermal Balance}

   In a schematic form, the equation of heat balance is
\beq
\sum_i \Lambda_i = \dot H,
\eneq
where each separate cooling mechanism is denoted by $\Lambda_i$.  The
processes  we include are:
\begin{description}
\item a)~~Bremsstrahlung  cooling, $\Lambda_B$, which is the sum of
the rates for $e-p$, $e-e$, and $e^+e^-$ interactions; explicit expressions
are given in BS91. Here we must mention that
there is
one more mechanism of internal photon production, the double
Compton process ($\gamma + e \rightarrow
\gamma+\gamma+e$). This can be important in the ``low''-temperature
range ($\Theta \ltwid 0.1$), in terms of number density of generated
photons (S84, PK94). On the other hand, from the energetics
point of view, it is negligible with respect to other
cooling mechanisms, since the created photons are generally at a much lower
energy than the original interacting photon, usually one of the
Comptonized  high-energy photons (see PK94, Thorne 1981)

\item b)~~Annihilation losses, defined here as $\Lambda_A = \Lambda_a -
 2m_e c^2 \dot n_A$, where $\Lambda_a$
is the total annihilation cooling rate;  this definition
  of the cooling rate accounts for the fact that we are interested in the
  losses to the thermal energy of the plasma, and the rest mass
  energy loss contribution is actually balanced by pair creation, since
  pair density is in equilibrium.
\item c)~~Inverse Compton losses on all (\ie bremsstrahlung and, possibly,
Double
Compton) internally created photons, written as
$$
(\Lambda_c)_{br} = m_ec^2 3\Theta n_{\gamma}/t_{esc},
$$
where $n_{\gamma}$ is given by eq.~(13).

\item d)~~Finally, Inverse Comptonization of the external soft photons,
including formation of both the Wien component and the power law component:
\begin{description}
\item{1)}~~The Compton cooling term due to the contribution of external
soft photons to the Wien peak is
\beq
(\Lambda_c)_{sw} =  m_ec^2 3\Theta n_{0w}/t_{esc},
\eneq
\item{2)}~~The Compton cooling due to the power law portion can be evaluated as
\beq
(\Lambda_c)_{pl}=  {m_e c^2 \over t_{esc}} \int_{x_0}^{\infty} n_{pl}(x) x
dx=m_ec^2{n_{0p}\over 2 t_{esc}}
	   \Theta \int_{x_0/\Theta}^{\infty} t^{-\alpha} e^{-t} dt=
	   m_e c^2{n_{0p}\over 2 t_{esc}}\Theta I_0(\alpha, x_0/\Theta),
\eneq
where
\beq
I_0(\alpha,x_0/\Theta)= \int_{x_0/\Theta}^{\infty}t^{-\alpha}
	 e^{-t} dt\simeq \cases { \Gamma(1-\alpha)& for $0 < \alpha < 1$\cr
	 (x_0/\Theta)^{-1/2}[|\ln (x_0/\Theta)|-\psi(1)]& for $\alpha=1$\cr
				(x_0/\Theta)^{(1-\alpha)}/(\alpha-1)& for $1< \alpha <2$\cr}.
\eneq
In this expression $\psi(z)$ is the logarithmic derivative of the gamma
function, and $\psi(1)= 0.577156649$.
\end{description}

  The escape time $t_{esc}$ is given by eq.~(14).

  One final correction is necessary: the preceding expressions actually
give the total emergent spectrum, which sums the inverse Compton cooling
rate with the luminosity injected in soft photons.
The net Compton cooling due to processing of the external seed photons
is therefore
\beq
(\Lambda_c)_s = (\Lambda_c)_{sw} + (\Lambda_c)_{pl} - {L_s
		\over 4\pi R^3/3}[1-exp(-\tau_T)].
\eneq

\end{description}

\section{Results}

\subsection{Independence with respect to $x_0$}

In the following subsections all the cases we illustrate were computed
with the same initial soft photon energy,
namely $x_0=10^{-5}$.  To check the sensitivity of our results to the
choice of $x_0$,
we have also computed equilibria with $x_0$ a factor of 10 larger and smaller.
On the basis of these results, we can assert that, provided
the basic condition $\Theta/x_0 \gg 1$ is fulfilled, for a fixed value of
$l_s/l_h$ there is only a very weak dependence of the equilibrium solutions
on the actual value of $x_0$.
For an order of magnitude change in $x_0$,
the corresponding variations in the equilibrium values of
$\Theta$ and $\tau_T$  typically amount to only a few per cent and
are never larger than $\sim 15\%$.  We are therefore confident that
our results are, indeed, representative of the general problem.

\subsection{Pair equilibrium curves: the effects of changing $n_0/n_p$}

   We begin by presenting a global view of our equilibria, with an emphasis
on illustrating how a variable soft photon supply can affect them.  In
order to facilitate comparison with previous work (\eg S84, BS91, PK94),
in which $\Theta$ was
generally taken as the independent variable, Fig.~1 displays the normalized
pair density $z = n_+/n_p$, the total optical depth $\tau_T$, the required
heating rate (as a compactness), and the spectral index $\alpha$ of the output
Comptonized power law component of the spectrum, all as functions of $\Theta$.
These equilibria share the same $\tau_p$ ($ = 1$) and used our standard value
of
$x_0$, $10^{-5}$.  In each panel we show a family of curves, parameterized
by $n_0/n_p$.  We remind the reader that fixed $n_0/n_p$ is equivalent
(for constant $\tau_p x_0$) to fixed $l_s$.

   Increasing $n_0/n_p$ clearly has a number of dramatic effects
on the character of the equilibrium:
\begin{description}
\item a)~~The maximum temperature, $\Theta_{max}$, that
       the system can attain steadily decreases; indeed, in almost all
       cases, the temperature that can be achieved for a given heating
       rate $l_h$ decreases as $l_s$ increases.
\item b)~~At fixed temperature, the pair content increases on the low-$z$
       branch, and decreases on the high-$z$ branch.
\item c)~~As corollaries of the previous point, at fixed temperature
       on the high-$z$ branch $\tau_T$ decreases and $\alpha$ increases,
       while the behavior is precisely opposite on the low-$z$ branch.
\item d)~~Again for fixed $\Theta$, on the low-$z$ branch increasing $n_0/n_p$
	 implies an increase in $l_h$; however, when $z > 1$, $l_h$ becomes
         almost independent of $n_0/n_p$ for fixed $\Theta$ and the different
         curves $l_h(\Theta)$ of Fig. 1c all converge.
\item e)~~At very large $z$ (and therefore $l_h$), the $z(\Theta)$ curves
          become independent of $l_s$.  The value of $l_h$
         at which this happens
         increases linearly with $n_0/n_p$; that is, the curves converge
         at a single value of $l_s/l_h$.
\end{description}
We now briefly elaborate on these points in order; most will be re-discussed
from the point of view of $l_h$ as the independent parameter in the following
subsection, and the physics behind them will then become much clearer.

   That the temperature falls as $n_0/n_p$ increases is no surprise; the
number of photons available for Compton cooling increases, forcing the
temperature to decrease.  Note that the {\it range} of temperatures
permitted also narrows as $n_0/n_p$ increases.

   Similarly, when there are many soft photons available, fewer electrons
are required to account for the total cooling at fixed temperature, so
the equilibrium pair content falls on the high-$z$ branch.  On the other
hand, there is a larger number of photons which may potentially be scattered to
energies above the pair production threshold, so the number of pairs
on the low-$z$ branch increases.

   Fewer pairs translates directly into smaller $\tau_T$, and smaller
$\tau_T$ implies less Comptonization.  Therefore, increasing $n_0/n_p$
spells a softer power-law on the high-$z$ branch, and a (very slightly)
harder one on the low-$z$ branch.

   The increase in $l_h$ at fixed $\Theta$ on the low-$z$ branch noted
in point d) is a corollary of point a).  Similar reasoning explains a
related effect,
not illustrated in Fig.~1: increasing $\tau_p$ decreases $\Theta_{max}$
at fixed $n_0/n_p$.  With a fixed number of seed photons, but greater
opportunity to scatter them, the temperature falls.  For example,
with $n_0/n_p = 10^5$, $\Theta_{max}$ falls from 1.12 for $\tau_p = 0.1$
to 0.216 for $\tau_p = 1$.

   The insensitivity of the $l_h (\Theta)$ curves to $l_s$ on the high-$z$
branch is due
to a combination of different effects.  For fixed $l_s$, the limit of very
large $l_h$ produces a condition in which $z \gg 1$, so that
$\tau_T$ becomes very large and Comptonization is saturated.
The output spectrum is then dominated, of course, by the Wien portion.  In that
limit, the ratio between Wien photons and particles in the plasma is
a function only of temperature because pair balance determines the number
density of photons.  When this is the case $l_h$ also depends only on $\Theta$
(S84).
%
%

At values of $l_h$ slightly lower than those at which complete Wien equilibrium
is attained (i.e., for slightly
higher temperatures on the high-$z$ branch), at fixed $\Theta$,
larger $l_s$ reduces $z$ and hence $\tau_T$ (on the low-$z$ branch, $l_s$
has little influence on $\tau_T$, of course).  This in turn leads to a
reduction in the amount of energy carried off by the average escaping photon
because Comptonization is weakened.  However, this reduction is almost
exactly compensated by the increase in the number of photons available due
to the increase in $l_s$.  We will be able to describe this phenomenon
more clearly and quantitatively armed with the formalism and scaling laws
developed in \S 4.4.

However, as Fig.~1 shows, when $l_s \gtwid 3$, this
compensation is no longer effective.  In this case, the compactness curve
crosses the bundle of curves describing smaller $l_s$ equilibria,
and does not join them until true Wien equilibrium is reached at
substantially larger values of $l_h$ (see also Fig. 1 in
Svensson 1986 for another illustration of this effect).
When $l_s$ is this large,
the greater number of photons available for Comptonization becomes
more important than the reduction in the mean energy per photon,
and a larger heating rate is required to maintain thermal balance.
Put another way, more energy is required to approach Wien equilibrium
because there are more photons to scatter up to energies $\sim kT$.

  Point e) is due to the fact that when $z \gg 1$, the increase in total
$e^{\pm}$ density leads to a large increase in the internal photon production
rate by bremsstrahlung.  At sufficiently large $z$, internal photon production
dominates over external soft photon input and the Compton cooling on these
internally generated photons becomes, in turn, comparable to or dominant
over other cooling contributions.  At this point the hard compactness is
basically determined by the ``internal'' conditions, i.e., it is (almost)
independent of the external photon input.
In Fig.~2 we show an example of this behavior, for the case $(\tau_p=1, n_0/n_p
= 10^4)$.  The various
cooling rates $\Lambda_i$ are plotted in the form of compactnesses:
\beq
 l_i = {4\pi\over 3}\tau_p{\Lambda_i\over m_ec^2n_p (c/R)},
\eneq
where the subscript $i$ indicates the specific process.  As Fig. 2 shows,
over most of the range of $l_h$, inverse Compton cooling on externally-created
photons accounts
for nearly all the cooling, but at the highest values of $l_h$, it
is overtaken by inverse Compton cooling on internal photons.
An inspection of the cooling rates for other cases confirms that
the onset of this regime (and the consequent convergence of the
$z(\Theta)$ curves) corresponds to the condition in which
the compactness ratio $l_s/l_h$ reaches a specific (rather extreme)
value, namely $\simeq 10^{-4}$.

\subsection{Plasma Temperature as a Function of $l_h$, $l_s/l_h$, and $\tau_p$}

  As we have just seen, discussing our equilibria in the conventional
language using $\Theta$ as the independent parameter makes many effects
difficult to explain because they require arguments based on self-consistency
that appear circular.  Much greater clarity is achieved by transforming
to $l_h$ as independent variable because fundamentally, the temperature
is the result of equilibration between a given heating rate and the various
cooling mechanisms.  Similarly, while $\tau_T$ is more closely connected
to the degree of Comptonization than $\tau_p$, the extra number of electrons
and positrons which changes $\tau_p$ to $\tau_T$ is a consequence of
the pair equilibrium, while $\tau_p$ is (we presume) controlled by the forces
acting on the plasma.  Finally, for the remainder of this paper we
will use $l_s/l_h$ as the third independent variable because it is more
directly related to the global energy budget (see \S 4.4) and the
geometry of the plasma than is $n_0/n_p$:
\begin{equation}
{l_s \over l_h} \simeq \left({L_{s,intr} \over L_h} + C\right)
{R^2 \over d^2} \phi ,
\end{equation}
where $L_{s,intr}$ is the intrinsic luminosity of the source of soft
photons, $C$ is the fraction of solid angle around the hot plasma occupied
by optically thick matter capable of absorbing the X-rays and re-emitting
their energy in lower energy photons, $d$ is the typical distance from
the soft photon source to the hot plasma, and $\phi$ accounts for possible
transfer effects within the hot plasma.  Because $l_s/l_h$ has such
a simple physical interpretation, it has been widely utilized elsewhere
(\eg Haardt \& Maraschi 1991, 1993; Ghisellini \& Haardt 1994; Dermer \etal
1991).

  In this section, we present the results of our calculations in terms of
these three independent parameters, and describe the trends we find.  In
the following section we will explain these trends physically.
The results we show here span a large volume of parameter space, but with
the greatest dynamic range in $l_h$: $0.03 \leq l_h \leq 3 \times 10^3$;
$0.01 \leq l_s/l_h \leq 1$; and $0.1 \leq \tau_p \leq 3$.
(We have actually computed equilibria for $l_s/l_h = 0.001$ as well, and
the corresponding results are included in the figures referred to
in \S~4.4.3 and in the related discussion.) This range
of parameters should cover most cases relevant to the study of accreting
black holes (see \S 5 for more discussion of this point).
We chose
$(\tau_p)_{max}=3$, because, with increasing $l_s$, higher values  of
$\tau_p$ would yield temperatures which fall below our range of interest
(for example, for $\tau_p=3$ and $l_s\simeq 0.94$,
$\Theta_{max}\simeq 0.11$).

In Figs.~3,4,5 the equilibrium curves as functions of $l_h$ are shown
for $l_s/l_h =1.0,0.1,0.01$ respectively; in each figure we plot
$\Theta(l_h)$, together with the curves for $\tau_T(l_h)$, $\alpha(l_h)$,
and $y_r(l_h)$ (as defined in eq.~(9)).  Each quantity is also
computed for several values of $\tau_p$.
In Fig.~3 ($l_s/l_h = 1$), $\tau_p =0.1, 0.5, 1.0, 2.0$; in Figs.~4
and 5, $\tau_p=0.1,0.5,1.0,3.0$.

The most prominent feature of these plots is the wide region in
$l_h$ over which, taking $\tau_p$ and $l_s/l_h$ fixed,
$\Theta$ is very nearly constant.  When $l_s/l_h \gtwid 0.01$ and $\tau_p
> 0.1$, the temperature changes by no more than a few percent over as much
as four decades in heating rate!

The existence of this constant temperature regime is not new; it is implicitly
contained in the results of Haardt
\& Maraschi (1991).  It appears (see \S 4.4 for a fuller discussion)
whenever inverse Compton scattering
of externally supplied photons is the dominant cooling mechanism, $l_s/l_h$
is fixed (as, for example, by absorption and subsequent reradiation in
soft photons of a fixed fraction of the hot plasma's radiative output),
and there are few pairs, so that $\tau_T \simeq \tau_p$.  The new points
we make here are to define the range of conditions in which it
applies, and to elucidate the physics which controls it.

The actual value of $\Theta$ on the flat part of the curve is also
rather insensitive to $l_s/l_h$: an order of magnitude increase in that
quantity decreases the temperature by only a factor of 2 -- 2.5.  It is,
however, more sensitive to changes in $\tau_p$: in the range of values
we have explored, $\partial \ln \Theta /\partial \ln \tau_p \simeq -1$
if $l_h$ is fixed somewhere in the range for which the temperature is
constant (Haardt \& Maraschi 1993 find a similar dependence of $\Theta$
on $\tau_p$ for the specific case of $l_s/l_h \simeq 0.5$).

The dynamic range in $l_h$ corresponding to
constant temperature also depends
on both the compactness ratio $l_s/l_h$ and $\tau_p$.  For
fixed $l_s/l_h$, larger $\tau_p$
yields a wider range of $l_h$ in which the temperature is constant.
Likewise, for fixed $\tau_p$, increasing $l_s/l_h$ also leads to a
widening of the interval in hard compactness over which $\Theta$ is constant.

The upper boundary in $l_h$ for the constant temperature range can
be anywhere from $\sim 1$ to $ > 3 \times 10^3$.  It is set by the point at
which
pair density starts to be no longer negligible ($z \gtwid 0.05$), so that
$\tau_T$ begins
to become greater than $\tau_p$, and this is sensitive to both $l_s/l_h$
and $\tau_p$.  In those cases where $\Theta$ remains
flat up to the largest value of $l_h$ shown, that is because $z$ remains
 very small throughout that range of $l_h$.

  The position of the lower boundary varies rather less than the position
of the upper boundary.  Throughout our parameter range it is $\sim 0.01$ --
0.1.
As we shall show in detail in the following subsection, the lower boundary is
set by the point at which
other cooling processes (notably bremsstrahlung) become competitive
with inverse Compton cooling.  The position of this boundary depends
hardly at all on $l_s/l_h$, but does depend somewhat on $\tau_p$: increasing
$\tau_p$ increases $l_h$ at the lower boundary.

At the very highest values
of $l_h$, all the equilibrium curves
corresponding to different values of $\tau_p$, for a given
ratio $\l_s/l_h$, tend to coincide,
again going towards Wien equilibrium conditions for the plasma.
This is clearly seen in Fig.~5,
corresponding to the smallest $l_s/l_h$ shown here.  This is in fact
another manifestation of the independence of $\Theta$ with respect to
$l_h$ when $\tau_T$ is fixed; it is just that when $l_h$ is very large,
$z$ is also very large, and $\tau_T$ now depends on $l_h$ rather than on
$\tau_p$. In fact, if we had plotted
these results for fixed $\tau_T$, the constant temperature range would
extend to even greater $l_h$, and the different curves would remain
separate even at very large $l_h$.  Note that the larger the value of
$l_s/l_h$, the larger the value of $l_h$ at which this occurs.  This is because
larger $l_s/l_h$ leads to lower temperature, and consequently, for
fixed $l_h$, a lower rate of pair production.

  We close this section by pointing out that, for fixed $l_s/l_h$, $\alpha$ is
independent of $l_h$ over an even broader range of $l_h$ than $\Theta$ is;
in addition, it is rather less sensitive to $\tau_T$.  Several factors
contribute to this.  In the range of $l_h$ for which $\Theta$ is constant,
$\tau_T$ is also constant, so $\alpha$, which is a function of those
two variables (\S 3.1.1), must certainly be constant.  However, at
larger $l_h$, where $\tau_T$ grows with the greater importance of pairs,
$\Theta$ falls with increasing $l_h$ in such a way as to cancel the effects of
changing $\tau_T$ (see \S 4.4.3).  Similarly, the weak dependence on $\tau_p$
in the low-$z$ regime is due to the inverse relation of $\Theta$ and $\tau_p$
(also elaborated more fully in \S 4.4.3), while the weak dependence on
$\tau_p$ in the high-$z$ regime is due to the decoupling of $\tau_T$ from
$\tau_p$ when $z \gg 1$.  The net result is that, provided $l_h$ is
greater than the lower boundary of the constant temperature range,
$\alpha$ is very nearly a function of $l_s/l_h$ alone.

\subsection{Cooling Physics}

\subsubsection{Relative importance of specific mechanisms}

  To explain the trends described in the previous subsection, we now
examine the relative importance of the different cooling mechanisms.
To illustrate this, we plot the ratios of
the different cooling compactnesses, $l_i$, to the total hard
compactness (i.e., the heating rate compactness) $l_h$ as
functions of $l_h$ itself for the same values of
$l_s/l_h$ and $\tau_p$ shown in Figs.~3, 4, and 5.
Thus, Figs.~6, 7, and 8 show these plots
for $l_s/l_h = 1.0, 0.1, 0.01$ respectively, with each of the
four panels corresponding to a given value of $\tau_p$.
The curves shown in each plot are: the ratio $l_{hc}/l_h$ of the total Compton
cooling compactness to the total compactness; the ratio $l_B/l_h$ of the
bremsstrahlung cooling compactness to the total compactness;
and the ratio $l_A/l_h$, where $l_A$ is the annihilation
cooling compactness.

It is apparent that Comptonization losses, $l_{hc}$,  including
both the contribution due to inverse Compton on the
internal bremsstrahlung photons and that of inverse Compton
on the externally-supplied soft photons
(that is,
\beq
l_{hc} = {4\pi\over 3}{\tau_p\over m_ec^2n_p(c/R)}[(\Lambda_c)_{br}+
	(\Lambda_c)_s],
\eneq
with the notations of section~3.2),
completely dominate over a large range in total hard compactness.
Compton cooling dominates all the way from the range where $l_h$ is so large
that pair processes dominate down to the bottom of the constant temperature
range.

Ultimately, at sufficiently small $l_h$, bremsstrahlung overcomes
inverse Compton cooling.  The value of $l_h$ where this occurs (and
therefore the lower boundary to the constant temperature range),
$l_{h,min}$, may
be estimated simply by comparing the bremsstrahlung cooling rate to
the total heating rate; we find
\beq
l_{h,min} \simeq c_o \tau_p^2 \Theta^q
\eneq
where $c_o \simeq 0.031$ and $q = 1/2$ when electron-proton bremsstrahlung
dominates, and $c_o \simeq 0.16$ and $q = 1$ when  electron-electron
bremsstrahlung dominates. These estimates are confirmed by the more
careful calculations shown in Figs. 6, 7, and 8.

Where $l_h < l_{h,min}$, we expect $\Theta \propto l_h^{1/q}$.  The curves in
Figs.~3, 4 and 5 allow us to qualitatively check
this behavior.  Because $q$ is itself an increasing function of $\Theta$
for the interesting range of temperatures, we expect $\Theta$ to
increase most steeply with $l_h$ in the bremsstrahlung-dominated regime
when $\Theta$ is least.  That occurs for the larger values of $\tau_p$
and $l_s/l_h$, an expectation which is at least qualitatively borne
out in these figures.

We also observe, comparing the cooling rate plots with the corresponding
temperature and optical depth equilibrium figures, that $l_A$ becomes
greater than $l_B$ at very close to the same place at which $z$
becomes $ > 1$.

\subsubsection{Thermal equilibrium in the inverse Compton regime: the
luminosity enhancement factor}

We have now identified the range in hard compactness
corresponding to conditions in which thermal Comptonization controls the
cooling of the system. In this case,  we have
$$
l_{hc} = l_h,
$$
and the considerations discussed in \S 2.2 should in principle apply.
Except at the very highest $l_h$, wherever inverse Compton cooling
dominates, externally-supplied photons dominate the internally-created.
In this section we will discuss how one might attempt to analytically
define scaling relations
between $l_s$, $l_h$, $\Theta$, and $\tau_T$ in this regime.

When inverse Compton on externally-supplied photons dominates,
we can evaluate the hard luminosity, $L_{h}$, as
\beq
L_h = {4\pi R^3\over 3}{m_ec^2\over t_{esc}}\int n_{sc}(x)xdx
-L_s(1-e^{-\tau_T}),
\eneq
where the second term is subtracted in order to obtain the net
Compton luminosity,  and $n_{sc}(x)$ is given by eq.~(15).  We define
\beq
<x>_1 \equiv {\int n_{sc}(x) x dx\over \int n_{sc}(x) dx},
\eneq
where the suffix 1 means that, with this definition (see eqs.~(15) and
(20)), the average over
photon energy includes only those photons which scattered
at least once.  These definitions yield a simple
relation between the input and output compactnesses, which is valid
when the spectrum of injected photons is much narrower than the output
spectrum (an assumption well-satisfied here):
\beq
l_{h}/l_s = \left[ {4\over 3}{\langle{x\over x_0}\rangle_1 \over
			 (1+g_{\tau}\tau_T/3)}-1\right](1 -e^{-\tau_T}),
\eneq
where $\langle x/x_0 \rangle_1$ is the mean factor by which the energy
of incoming photons is amplified before escape, the factor
$1 + g_\tau \tau_T/3$
is the ratio between the escape time for the photons
and the time $R/c$, and the factor of $e^{-\tau_T}$ corrects for those soft
photons which pass through without scattering at all.

As in Dermer \etal (1991), the compactness ratio can be formally expressed
in terms of the luminosity enhancement
factor, $\eta$:
\beq
L_{hc} = {4\pi R^3\over 3} m_ec^2 \int x'\dot n_{soft}(x') [\eta(x',
\Theta,\tau_T)-1],
\eneq
where
$\dot n_{soft}(x')$ is the spectral soft photon injection rate,
and $\eta$ is in general a function of the initial soft photon energy
$x'$, in addition to its dependence on $\Theta$ and $\tau_T$.  In our case
$\dot n_{soft}(x') = n_0\delta(x'-x_0)/(R/c)$.  The integrand is
multiplied by $\eta - 1$ rather than $\eta$ so that $L_{hc}$ is
the net Compton cooling.  Consequently, for the present problem,
from eq.~(35), we have
\beq
{L_{h}\over L_s} = {l_{h}\over l_s} = \eta(x_0, \Theta, \tau_T) -1,
\eneq
and, comparing with eq.~(34),
\beq
\eta \equiv e^{-\tau_T} + {4 \over 3} {\langle {x \over x_0}\rangle_1 \over
1 + g_\tau \tau_T/3} \left(1 - e^{-\tau_T}\right).
\eneq

One commonly followed approach to computing $\eta$ is to adopt a simple model
of
inverse Compton scattering in which to reach a given amplification $x/x_0$
always requires exactly $n(x/x_0)$ scatterings.  If $x \ll \Theta$,
$n(x/x_0) \simeq \ln(x/x_0)/\ln(A_r)$ ($A_r$ is defined in equation (7)).
In terms of this model,
\beq
\langle {x \over x_0} \rangle = \sum_0^{\infty} P_n  {x \over x_0} (n),
\eneq
where $P_n$ is the probability of escape after exactly $n$ scatterings.
Note that the relation between $\langle {x / x_0} \rangle$, accounting for
those
photons that are not scattered at all, and the average we have
introduced above (eq.~(33)), $\langle {x / x_0} \rangle_1$, is simply
\beq
\langle{x\over x_0}\rangle = \langle{x\over x_0}\rangle_1 (1-e^{-\tau_T}) +
							  e^{-\tau_T}.
\eneq

Several different approximation schemes have been proposed to estimate
this sum.  The simplest (\eg Zdziarski \etal 1990) effectively replaces
$P_n$ with
$\delta_{nm}$, where $\delta_{ij}$ is the Kronecker delta, and $m$ is the mean
number of scatterings before escape.  Loeb \etal (1991) suggested
approximating $P_n$ by the Poisson distribution.  For the problem at hand,
however, $P_n$ is best described by the solution of the ``gambler's ruin"
problem (see, {\it e.g.} Feller 1967).  The high-$n$ tail of this distribution
falls rather more slowly than in the Poisson distribution.  At the same time,
the amplification $A_r^n$ increases exponentially with $n$.  Therefore, the
largest contribution to the sum in equation
(38) is likely to come from a range of $n$ rather larger than $m$,
and the associated amplification can be larger by a very substantial factor.
To check just how bad an approximation the Poisson distribution is, we have
compared the predictions of the Loeb \etal model (using $y_r$ as given by
eq.~(9)) to the actual values
of $\eta(\Theta,\tau_T)$ found in our solutions.  The ratio between the
two exhibits a very large scatter, with a general offset in the sense that
the Poisson model substantially underestimates the true amplification.

  For these reasons, an alternative description of Comptonization was
proposed by Dermer \etal (1991).  They  basically suggested that the
amplification
factor $\langle x/x_0\rangle$ should be computed directly from the output
spectrum (which they evaluated with a Monte Carlo simulation).
We approximate the  output spectrum analytically by a power-law plus a
Wien component, so we write
\beq
\langle {x \over x_0} \rangle_1 = f_{pl}{\langle x_{pl} \rangle \over x_0}
+ f_W {\langle x_{W} \rangle \over x_0}
\eneq
where $f_{pl},f_{W}$ are the fractions of the scattered photons going into the
power-law and the Wien component, respectively, and $f_{pl} + f_W = 1$.
The power-law fraction is
\beq
f_{pl}( x_0,\Theta, \tau_T)  \equiv
          {\int_{x_0}^{\infty} n_{pl}(x) dx
		   \over n_0(1-e^{-\tau_T})} =
{1\over 1+ 2\alpha{\Gamma(\alpha)
         \over \Gamma(2\alpha+3)}P_{\tau_T}(\tau_T)\left ({x_0
         \over \Theta}\right )^{\alpha}},
\eneq
where we have used equations~(19) and (26).  The average photon energies are
\beq
<x_{pl}> = \cases{ x_0\alpha\left ({\Theta\over x_0}\right )^{1-\alpha}
           \Gamma(1-\alpha)
        & for  $0< \alpha < 1$\cr
           x_0{\alpha\over (\alpha-1)} & for $ \alpha >1$\cr},
\eneq
(here we neglect to specify the case for $\alpha =1$ exactly);
and $\langle x_W \rangle = 3\Theta$.

  In fact, over almost the entire range of $l_s/l_h$ we have studied,
$f_{pl}\simeq 1$, i.e., $f_w << f_{pl}$.
$f_w$ increases with $l_{hc}/l_s$, but the greatest value we find for
it is $\ltwid 4\times 10^{-2}$ for $l_{hc}/l_s = 10^3$.  At this
extreme value, the energy in the Wien term begins to be competitive with
the energy in the power-law term; everywhere else it is negligible.  Thus,
the power-law portion of the inverse Compton losses is nearly always the
dominant one.

   We therefore make the approximations that
$f_{pl} \simeq 1$ and $f_{pl}\langle x_{pl}\rangle \gg f_W \langle x_W
\rangle$.
In this limit, when $0 < \alpha < 1$ (nearly all our equilibria fall in
this range),
\beq
{l_h \over l_s} \simeq \left(1 - e^{-\tau_T}\right)\left[{(4/3)\alpha
\Gamma(1-\alpha) (\Theta/x_o)^{1-\alpha} \over
1 + g_{\tau_T} \tau_T/3} - 1\right].
\eneq
Because $\alpha$ depends in a rather complicated way on $\Theta$ and $\tau_T$,
there appears to be no simple analytic relation describing the thermal
balance in this regime.

   On the other hand, when $\alpha > 1$, the analogous relation becomes
$${l_h \over l_s} \simeq \left(1 - e^{-\tau_T}\right)
\left[{(4/3)\alpha/(\alpha - 1) \over 1 + g_{\tau_T}\tau_T/3} - 1\right].$$
Once again, the absence of a simple functional relation between $\alpha$
and $\Theta$ and $\tau_T$ prevents further reduction of this form.

\subsubsection{Empirical scaling}

Our numerical results do, however, reveal several simple scaling laws.
Fig.~9 shows $\Theta$ as a
function of the corresponding equilibrium $\tau_T$, for all our
equilibria for which inverse Compton cooling dominates. It is apparent that the
points for each value of $l_s/l_h$ describe very nearly a straight line
in this diagram, and these lines are all very nearly parallel, with
logarithmic slope $\simeq -1$. There is no break apparent in any of these
lines, at $\tau_T = 1$ or anywhere else.  For fixed $\tau_p$, the spread
in $\tau_T$ is caused by increasing numbers of pairs.  At the same time,
differing values of $\tau_p$ which share the same $l_s/l_h$ and happen
to yield the same $\tau_T$ also coincide in $\Theta$, demonstrating that
$\tau_T$ is the appropriate physical variable for this correlation.

We are thus led to plot, in Fig.~10a, the product $(\Theta\tau_T)$
as a function of the ratio $l_s/l_{hc}$, for the same choice of
equilibrium solutions; Fig.~10b shows
the corresponding values for $\alpha$, again versus $l_s/l_{hc}$.
Again, the relationships are very nearly linear, and with very little
scatter.  Fitting to the points in Fig.~10a yields the relation
\beq
l_{hc}/l_s \simeq (10\Theta\tau_T)^{\beta},
\eneq
where $\beta = 4 \pm 0.5$.  This relationship is a much better
description of Comptonization equilibrium than any expression of the
form $e^y$. This
empirical scaling law is, of course, what lies behind the trends we
described in \S 4.3.

The physical origin of the correlation with $l_s/l_h$ can be easily seen,
at least
qualitatively.  The cooling rate depends on the seed photon energy
density, which rises with increasing $l_s/l_h$.  However, because
even unsaturated Comptonization can significantly increase the photon
energy density
above the energy density directly injected, there is a nonlinear
dependence on the two quantities which promote Comptonization, namely
$\Theta$ and $\tau_T$.  That the nonlinear dependence should be
describable so simply is surprising, given the arguments of
the preceding section, but it does appear to be correct.

This scaling law is not completely new.  Haardt \& Maraschi (1991)
demonstrated that $l_s/l_h$ fixes a function of $\Theta$ and $\tau_T$, and in
Haardt \& Maraschi (1993) they presented results showing that when $\tau_T <
1$,
this function depends approximately on the product
$\Theta\tau_T$ rather than on the two variables separately (at least for
$\Theta$ not too large).  What is new here is two things:
the extension of this relation to $\tau_T > 1$, and the scaling
with $l_s/l_h$.

   In addition, we find that the coefficient in this scaling law
depends weakly on geometry.
For fixed $l_s$ and $\tau_T$, a plane-parallel
slab has a higher average soft photon energy density than a sphere
because the mean number of scatterings to escape is greater.  Greater
photon energy density leads to stronger cooling, and hence, for fixed
$l_h$, a lower equilibrium temperature.  Consequently, the coefficient
of $\Theta\tau_T$ in equation (44) changes from 10 to $\simeq 16$.

We have already remarked (\S 4.3) that in the inverse Compton dominated
regime, $\alpha$ is essentially a function of $l_s/l_h$ alone.
In Fig.~10b the plot
shows that functional relation.  Fitting to these points yields
\beq
\alpha \simeq 1.6 \left({l_s \over l_h}\right)^{0.25}.
\eneq
Comparing the scaling laws
shown in the two parts of Fig.~10, we find the very simple relation
\beq
\alpha \simeq {0.16 \over \Theta \tau_T},
\eneq
as can be visually inferred from  Figs.~10a, and 10b.  The
coefficient 0.16 applies in the case of spherical geometry; in slab
geometry, it becomes 0.1.

When $l_s/l_h \ll 1$,
there is hardly any scatter around the relation given in equation 45;
however, when $l_s/l_h$ is more than a few tenths and $\tau_T \leq 1$, there
is also a weak secondary dependence on $\tau_T$ of the form $\alpha \simeq 1.4
\tau_T^{0.12}$.  For smaller $l_s/l_h$ or larger $\tau_T$, the spectral
index is essentially independent of $\tau_T$.

In slab geometry, the increased number of scatterings for fixed $\tau_T$
compensates for the lower temperature, so that the relation
between spectral index and $l_s/l_h$ is essentially identical to the
spherical case.  Slightly more energy is carried off in the Wien
component however, to make up for the lower energy cut-off in the power-law
component.

The scaling shown in equation 46 is in fact a fairly good approximation to the
expression for $\alpha(\Theta,\tau_T)$ given in
\S 3.1.1 in the case of modest optical depth ($1 < \tau_T < 3$) and low
temperature ($\Theta \ltwid 0.3$):
   $$ \alpha\sim {-\ln [1-3/(4\tau_T)]\over 4\Theta} \sim
		    {3\over 16 (\tau_T\Theta)}.$$
This expansion predicts the correct scaling and very nearly the correct
coefficient (0.19 instead of 0.16).  That the scaling law is in fact
more broadly applicable is due to the fact that we consider only the
portion of the $\Theta$ -- $\tau$ plane in which equilibria having
$l_s/l_h$ in the range $10^{-3}$ -- 1 are found.  This condition restricts
$\Theta \tau$ to be $\sim 0.1$, and when this condition is met, equation (46)
continues to be a fairly good approximation to the exact expression even
when $\tau < 1$.  That is, equation (46) only applies when
the plasma is in thermal balance.

   Just as for the temperature scaling law expressed in equation (44),
hints of this relation between spectral index and $l_s/l_h$ have
also appeared in the literature.  Gil'fanov \etal (1994) argued on a
qualitative basis that $\alpha$ should
depend primarily on $l_s/l_h$, and vary in the sense that we have found.
Working solely on the special case $l_s/l_h \simeq 0.5$,
Haardt \& Maraschi (1993) computed the spectral index
over the range $0.01 < \tau_T < 1$ and found that it hardly varied (the
sharp gradient in $\alpha$ they found as $\tau_T$ approaches 1 is due
to a breakdown in their approximations in this regime).
What we have done is to confirm that this relation exists, and to discover
its quantitative character.

  We have now arrived at a much more powerful description of the scalings
between physical parameters in the inverse Compton-dominated regime. It
is no accident that they have been found by a systmatic approach focusing
on causal parameters, rather than fitting to phenomenological parameters.
In fact, an approach which treated $\Theta$ and $\tau_T$ as independent
parameters would have been incapable of discovering these correlations
because there would have no sensible way to limit the region of the $\Theta$ --
$\tau_T$ plane considered.  Only framing the problem with $l_h$ and $l_s/l_h$
as the independent variables reveals these correlations.

\section{Consequences for Observable Systems}

\subsection{The Natural Ranges for $l_h$, $l_s/l_h$, and $\tau_p$}

\subsubsection{$l_h$}

   As we have just seen, the temperature is confined to a rather narrow
range over a very broad span of $l_h$ and $l_s/l_h$, provided $0.1 \ltwid
\tau_p \ltwid few$.  However, this volume of parameter space, large as it is,
is not infinite.  We now argue that it would be however, very unlikely for
any accreting black hole to be found outside this volume.

   First, as many have previously pointed out (\eg Lightman and Zdziarski
1987),
if the X-ray-emitting plasma is roughly spherical, $l_h$ may be rewritten in
terms of quantities more directly related to the accretion process:
\begin{equation}
l_h = {L_h \over L} {L \over L_E} {2\pi \mu_e/m_e \over r/r_g},
\end{equation}
where $L/L_E$ is the total luminosity relative to the Eddington luminosity,
$\mu_e \simeq 1.2 m_p$ is the mass per electron, and $r/r_g$ is the
size of the region relative to the gravitational radius of the central mass.
Most of the gravitational energy release in accretion onto a black hole
takes place around 5 -- $10r_g$, so it would be difficult to make $l_h$
more than $3 \times 10^3$.  On the other hand, if the portion of the
dissipation going into the X-ray-emitting plasma in that region is at least
$\sim 10^{-5} L_E$, $l_h$ is $ > 10^{-2}$, the lower limit of the
region in which we find almost constant temperature.  Thus, if there
is enough energy going into the X-ray-emitting plasma near a black hole
to make it detectable, $l_h$ will fall into the constant temperature range.

  In this respect, coronae near the accretion disks around weakly-magnetized
neutron stars should be very similar to those near black holes.  Accreting
white dwarfs generically should have
$l_h$ at least $\sim 10^{-5}$ times smaller for equal mass accretion rate,
and are therefore likely to fall just below the constant temperature range
of $l_h$ even if $L_h/L \simeq 1$.  With $l_h \sim 10^{-4}$ -- $10^{-3}$,
coron\ae~ in the inner regions of accretion disks around white dwarfs
are likely to be in a bremsstrahlung-dominated regime (see Fig.~6) with
temperatures of a few keV.

   The assumption of roughly spherical geometry is not entirely innocuous.
It is almost certainly valid for large $l_h$, where the plasma
is always pair-dominated.  When that is true, the electron thermal speeds
are comparable to the escape speed even for the smallest possible $r/r_g$, and
the positrons cancel out any restraining electric fields.  In addition,
the effective Eddington luminosity is reduced by $\sim 10^3$. However,
at the small $l_h$ end, where the plasma is not pair-dominated,
a less symmetrical geometry (\eg disk-like) can make a difference.
For fixed $\tau_T$, the ratio of bremsstrahlung luminosity to inverse
Compton luminosity is $\propto (r/h)^2 l_h^{-1}$, where $h$ is the
smaller dimension.
Thus, flattened geometry can cause the lower bound of the constant
temperature range of $l_h$ to increase substantially.

\subsubsection{$l_s/l_h$}

 If $L_{s,intr}/L_h \ll C$, $l_s/l_h$ is controlled by essentially
geometrical factors and is (almost) always $\leq 1$ (see equation 29).  A
common assumption is to suppose that the hot plasma forms a slab bounded on one
side by a (comparatively) cool accretion disk (Liang 1979, Haardt \&
Maraschi 1991).  In that case, $l_s/l_h \simeq
0.5$, with some order unity corrections due to anisotropy in the
upscattering (Ghisellini \etal 1991; Haardt 1993).  When the intrinsic
soft luminosity is small, while it is very difficult to make $l_s/l_h
> 1$, there are many ways it could be $\ll 1$.  Both $C$ and $r^2/d^2$,
for example, must be less than unity, and could be much less.  In
addition, modest relativistic motion can alter the effective value of
$l_s/l_h$ because this ratio should be evaluated in the bulk frame of
the hot plasma.  If the hot plasma moves away from the soft photon
source with speed $\beta$, the comoving energy density of soft
photons falls as $[\gamma (1 + \beta)]^{-4}$.  Motion towards the
soft photon source would have the opposite effect, of course, but it
would take rather contrived dynamics to produce such a situation.

   When $L_{s,intr}$ is not negligible, $l_s/l_h$ can easily be $ > 1$.
However, as shown in \S 4.4.3, the equilibrium temperature on the flat part
of the curve is $\propto
(l_s/l_h)^{-1/4}$, so rather large values of $l_s/l_h$ are necessary
in order to substantially cool the plasma.

\subsubsection{$\tau_p$}

  As is demonstrated in Figures 3, 4, and 5, the total optical depth $\tau_T$
becomes independent of $\tau_p$ (and a function only of $l_h$ and $l_s/l_h$)
when $l_h$ is large enough for the number of pairs to be much larger than
the net lepton number.  Regulation of $\tau_p$ is therefore only relevant
at smaller $l_h$.  The remainder of this subsection therefore focusses
on the range of parameters in which there are relatively few pairs
and $\tau_T \simeq \tau_p$.

    So far we have relied on a mental picture in which the hot X-ray-emitting
plasma is physically well separated from the cool source of soft photons.
If they actually touch each other (which is a statement almost equivalent
to $l_s/l_h \sim 1$), we must consider thermal mixing via
electron conduction.  Good magnetic connectivity is probably a reasonable
assumption because the most likely way to actually supply the energy to
dissipate in the hot plasma is through the passage of MHD waves from an
accretion disk to an adjacent hot corona (Stella and Rosner 1984; Tout and
Pringle 1992).  If wave-particle scattering does not grossly diminish
the hot electron mean free path, it makes little sense in this case to speak of
a hot plasma unless it is significantly thicker than a Coulomb mean free path.
If it were otherwise, much of the volume occupied by the hot electrons would
actually be shared with the cool matter (which would then be evaporating
into the hot gas: Balbus and McKee 1982).  This restriction places a lower
limit on $\tau_T$,
\begin{equation}
\tau_T \gg {8\pi \over 3} \Theta^b /\ln \Lambda ,
\end{equation}
where $\ln \Lambda$ is the usual Coulomb logarithm, and $b$ changes from 2
to 4 as $\Theta$ increases from $ \ll 1$ to $\gg 1$.  If $\Theta > 0.1$,
the lower limit on $\tau_T$ due to thermal conduction is $\sim 0.01$.

   When the optical depth becomes large, gradients in the radiation pressure
develop.  Large forces can be associated with these gradients; the
minimum radiation force is larger than the force due to gas pressure
gradients when $l_h > 4\pi \Theta$ in spherical
geometry.  In the presence of gravity, radiation pressure dominance often
leads to dynamical instabilities (\eg Arons, Klein,
\& Lea 1987; Krolik 1979).  We speculate that these dynamical instabilities may
place an upper limit on the optical depth by causing the hot plasma to
break up into smaller pieces whose individual optical depths are no
bigger than a few.

\subsection{The Connection between Spectral Index and Intrinsic Soft Luminosity
or Source Geometry}

    AGN exhibit a rather small range of power law indices in the X-ray band.
Type 1 Seyfert galaxies AGN have intrinsic
spectral slopes in the 2 -- 20 keV band of $\simeq 0.9$ with very little
scatter if the effects of Compton reflection
and partially-ionized absorption are factored in (Mushotzky \etal 1993).
Higher luminosity AGN (\ie quasars) have similar spectral slopes, although
possible contributions of reflection and ionized absorbers are harder
to determine (Mushotzky \etal 1993).

If the thermal Comptonization model is correct, we can infer that
$l_s/l_h$ is most likely held stably at $\sim 0.1$ in all of these AGN.
Such a small value of $l_s/l_h$ implies that at
least one of the following two conditions applies: both $L_{s,intr}/L_h$
and the covering fraction of any reprocessing surface must be small;
or the distance from the source of soft photons must be rather greater
than the size of the X-ray emitting region.  In type 1 Seyfert galaxies,
the frequent evidence for strong Compton reflection suggests $C \simeq 0.5$;
where this is the case, $(R/d)^2\phi$ must be $\sim 0.1$.  Note that
contributions to $C$ from the obscuring torus (Krolik \etal 1994; Ghisellini
\etal 1994) automatically have $(R/d)^2 \ll 1$.

  Moreover, because $\alpha$ may be used to infer $l_s/l_h$, and $l_s/l_h$
fixes $\Theta \tau_T$, a characteristic $\alpha \simeq 0.9$ implies that
$\Theta \tau_T \simeq 0.1$ -- 0.2. The smaller value
obtains when the plasma is a slab, the larger value when it is spherical.

  The only possible alternative to these inferences is a very specific
version of the
coronal model (\eg Haardt \& Maraschi 1991, 1993; Zdziarski \etal 1994),
in which $C \simeq 0.5$ and $R \simeq d$.  Because $l_s/l_h \simeq 0.5$ in
these models, they fall in the regime in which there is a small secondary
dependence of $\alpha$ on $\tau_T$.  For this value of $l_s/l_h$, the
observed spectral slope can be produced if $\tau_T$ is somewhat less than 0.1.
The uniformity of AGN spectra would then require a rather narrow
distribution of $\tau_T$.  It follows as a corollary that $\Theta$ would
also have a narrow distribution, centered on $\simeq 0.8$.
Heat conduction (\S 5.1.3) might pose a physical self-consistency problem for
models of this sort.

  If $l_s/l_h$ is really $\sim 0.1$,
because the observed ratio of $L_s/L_h$ is often order unity or somewhat
greater, some means must be found to reduce the {\it local} value of
$l_s/l_h$ in the the X-ray production region.  For example, it might not be
immediately adjacent to
the region where the bulk of the ultraviolet luminosity is radiated.
Another possibility is that the X-ray region moves away from the disk
fast enough that the seed photon intensity is diminished by redshifting.
All these schemes are made easier if the intrinsic soft luminosity is
minimized.

    In stellar black hole binaries, the X-ray spectrum is most simply described
as the sum of a quasi-thermal soft component and a hard power-law (Tanaka 1989,
White 1994).  The soft ({\it i.e.} $T \simeq 1$ -- 3 keV) component is
presumably due to dissipation in the accretion disk proper, where the density
and optical depth are great enough to approximately thermalize the spectrum.
Weakly-magnetized accreting neutron stars should also possess such a soft
component, although much of the dissipation may occur in shocks near the
surface of the neutron star rather than as a result of gradual viscous
dissipation in the disk proper.  In addition, accreting neutron stars may
have rather weaker hard components than accreting black holes
if the disk is interrupted more than a few neutron star radii away from
the surface.

  The hard power-law seen in accreting black hole systems is generally
thought to be the result of thermal Comptonization (Shapiro \etal 1976).
Spectral indices of this hard component exhibit a greater range
than in AGN; examples are known from $\alpha \simeq 0.3$ to
$\alpha \simeq 1.5$ (Tanaka 1989; Ballet \etal 1994; Gil'fanov \etal 1994), but
the center of the distribution is fairly close to the average spectral
index in AGN. We therefore expect a larger range in $l_s/l_h$,
but the mean inferred $l_s/l_h$ for stellar black holes is also $\sim
0.1$.  As for AGN, very optically thin coronal models can also produce
spectra with $\alpha \simeq 1$, but it may be difficult for them to generate
spectra as hard as $\alpha \simeq 0.3$.

\subsection{Coupling between Fluctuations in the Luminosity and the Spectrum}

    The simplest model for X-ray fluctuations one might imagine is one in which
the geometry and
the relative division of heating between hot plasma and cool gas (\ie the
factors which determine $l_s/l_h$) are fixed.  If the column density of the
corona is fixed
independently, changes in luminosity drive changes in $l_h$, but
not in the other parameters.  In the large volume of parameter space in
which the temperature varies only slowly with $l_h$, one would then
expect that
luminosity fluctuations should raise the general continuum level without
altering its spectrum, for $\Theta$ and $\tau_T$, which determine the
X-ray spectrum, are nearly invariant.

   In only one AGN (NGC 4151) are the effects of absorption and Compton
reflection weak enough that one can attempt to measure possible variations
in the intrinsic continuum shape (Yaqoob 1992).  In this object (whose
mean spectral index is comparatively small, $\simeq 0.5$), $\alpha$
rises from 0.3 to 0.7 while the 2 -- 10 keV flux grows by about a
factor of four.  In the framework of our model, this behavior would
be interpreted as an increase in $l_s/l_h$ from $\sim 0.001$ to $\sim 0.04$.
Whether this change in $l_s/l_h$ can be genuinely associated with a change
in total luminosity is hard to say because the total power in spectra this
hard is dominated by the highest energy photons, and the data which
exhibit this correlation say nothing about the flux at the photon energies
which dominate the output.

   There is a much greater wealth of observations regarding stellar
black hole X-ray variability than regarding AGN X-ray variability.
In the case of stellar black holes, the amplitude
of the hard component often fluctuates by very large amounts: changes of three
orders of magnitude have been seen.  Yet despite these very large fluctuations,
in most examples the spectral index has remained essentially fixed (Tanaka
1989;
Churazov \etal 1993; Gil'fanov \etal 1994).   If the {\it ratio} of soft
flux to hard were
constant, this behaviour would, of course, be precisely in keeping with the
simplest model one could imagine.

   However, the hard and soft components in fact appear to vary quite
independently.  Therefore, if the power law index of the hard component
remains constant despite changes in $L_s/L_h$, there must be an additional
element in the model beyond the simplest case.  One possibility is that
the main source of soft photons is quite distant from the radiation region
for the hard component, so that the soft photons serving as Comptonization
seeds are entirely from local reprocessing with constant reprocessing
fraction.  Another possibility is that the
distance between the Comptonizing plasma and the main source of soft photons
changes in such a way as to cancel the change in luminosity ratio.
For example, if the Comptonizing plasma is heated by dissipation of MHD
waves originating in an accretion disk, an increase in the intrinsic soft
luminosity of the disk might be accompanied by an increase in the flux of MHD
waves into the hotter plasma.  As a result, the hot plasma could be pushed
farther away from the disk, so that the soft photon density in the
Comptonizing region is reduced.

   In at least one example (Nova Muscae: Ebisawa \etal 1994), the spectral
index of the hard component {\it did} change.  In this case, as the soft
component faded, the spectral index of the hard component fell from $\simeq
1.5$ to $\simeq 0.7$.  At least in a qualitative sense, this follows
the general trend of decreasing $l_s/l_h$ leading to harder Comptonized
power-laws.

   The direct relation between spectral shape and $l_s/l_h$ breaks down
on the shortest timescales.  Spectral equilibration requires a time
$t_{eq} \sim \ln (\Theta/x_o) R/(c\tau_T)$ because the highest energy photons
are scattered $\simeq \ln (\Theta/x_o)$ times.  On the other hand,
$l_s/l_h$ could easily change in a time $R/c$, which is typically an
order of magnitude less than $t_{eq}$.  Therefore,
these correlations should not be followed on timescales shorter than $t_{eq}$.

\section{Summary}

    In this paper we have demonstrated that the prevalence of $\simeq 100$
keV thermal
Comptonization spectra in the X-ray emission from accreting black holes
is no coincidence.  Over a very wide range of dimensionless heating
rate (i.e. compactness $l_h$ ), and normalized strength of seed photon
supply (i.e. $l_s/l_h$), and a somewhat narrower range of net lepton
optical depth ($\tau_p$), thermal Compton scattering dominates the heat
balance, and the temperature is nearly always approximately of this
magnitude.  In fact, for fixed $l_s/l_h$ and $\tau_p$, the temperature
is virtually independent of heating rate over a dynamic range in $l_h$
that is generally at least several orders of magnitude.

   After quantifying the boundaries of this constant temperature regime,
we have also argued that nearly all accreting black hole systems should
fall within its (large) volume of parameter space.  That $l_h$ and $l_s/l_h$
should fall within that range seems very likely, given what we know of
black hole accretion dynamics; that $\tau_p$ should be in the range giving
this result is less certain, but we have raised several plausible arguments
toward this end.  The compactness is determined by the nature of accretion
into a relativistically deep potential; the seed photon supply follows
simply from geometrical considerations; the optical depth may be determined
by a combination of thermal conduction and radiation pressure dynamics.

   We have also shown that simple scaling laws relate the temperature,
total ({\it i.e.} including pairs) Thomson optical depth, and the seed
photon supply $l_s/l_h$.  $l_s/l_h$ (almost) uniquely determines the product
$\Theta \tau_T$, and, to roughly the same level of approximation,
$l_s/l_h$ alone determines the spectral index $\alpha$ of the
power-law segment of the Comptonized spectrum.

    On the basis of this relationship between $\alpha$ and $l_s/l_h$ we
have argued that $l_s/l_h$ in AGN is most likely to be
generically $\sim 0.1$.  Such a small value of $l_s/l_h$ requires any
thermalizing surface (for example,
an accretion disk) to be comparatively distant from the X-ray emission
region, and suggests that the intrinsic (as opposed to reprocessed) soft
photon luminosity is comparatively small.  If instead the thermalizing
surface is nearby, so that $l_s/l_h = 0.5$ or more, the optical depth of the
corona must be rather small, and its value is
quite tightly constrained; the coronal temperature is then similarly tightly
constrained.  In stellar black hole binaries, the observed range of power
law indices is larger than in AGN, so a larger range of $l_s/l_h$ is
permitted, but this range is also roughly centered on 0.1.  When stellar
black hole binaries make transitions between ``soft" and ``hard" states,
the slope of the Comptonized power law usually remains constant, even
while the relative amplitude of the soft thermal and hard Comptonized
components change by orders of magnitude.  This fact requires the
luminosity in seed photons fed into the Comptonizing plasma to not
be simply proportional to the soft luminosity we observe.

\acknowledgments

   We thank Francesco Haardt and Andrzej Zdziarski for many enlightening
conversations.  We also thank Andrzej Zdziarski for much guidance in the
nuances of pair equilibria and Comptonization.

   This work was partially supported by NASA Grant NAGW-3129.


\clearpage

\centerline{Figure Captions}

\vskip 0.4cm

\noindent Figure 1 \qquad A global view of our equilibria, seen as
functions of $\Theta $ for $\tau_p = 1.0$ and $x_0=10^{-5}$.  The
various curves are each for a different fixed value of $n_0/n_p$
(and, correspondingly, of $l_s$:
solid is $n_0/n_p = 0$; dotted is $n_0/n_p = 100$ (i.e., $l_s = 3.14\times
10^{-3}$); short dash is $n_0/n_p = 10^3$ ($l_s = 3.14\times 10^{-2}$);
long-dash is $n_0/n_p = 10^4$($l_s = 0.314$); dot-short-dash is $n_0/n_p =
10^5$
($l_s = 3.14$);
dot-long-dash is $n_0/n_p = 10^6$ ($l_s = 31.4$).
In the first three panels $l_h$ increases upwards; in the fourth it is
the opposite.

a) The normalized pair density $n_+/n_p$.  Pairs dominate only
when the equilibrium is well on the high-$l_h$ side of the temperature maximum,
but they can significantly influence the temperature even when $z \sim 0.1$.

b) The total Thomson optical depth $\tau_T$.

c) The heating rate in terms of compactness $l_h$.  The curves for different
$n_0/n_p$ converge when they become pair-dominated.

d) The power law index of the Comptonized output spectrum.  No curve is
plotted for $n_0/n_p = 0$ because this spectral component does not exist
in that limit.

\noindent Figure 2 \qquad The relative importance of different cooling
processes as functions of $l_h$ for the specific case $\tau_p = 1$,
$n_0/n_p = 10^4$ ($l_s = 0.3$).  The solid line is inverse Compton cooling
on external photons; the dotted curve is inverse Compton cooling
on internally created photons; the dashed curve is annihilation cooling;
and the dot-dashed curve is bremsstrahlung.  For most of the parameter
range, inverse Compton cooling on externally-created photons accounts
for nearly all the cooling, but at the highest values of $l_h$, it
is overtaken by inverse Compton cooling on internal photons.

\noindent Figure 3abcd \qquad The principal physical quantities as functions of
$l_h$ for $l_s/l_h = 1$.  In this figure and Figs.~4 and 5, we show
$\Theta$ (panel a), $\tau_T$ (panel b), $\alpha$ (panel c), and $y_r$ (panel d)
as functions of $l_h$.  In each of Figs.~3, 4, and 5, the solid line
corresponds to $\tau_p = 0.1$; the dotted line to $\tau_p = 0.5$; the short
dashed line to $\tau_p = 1$; the long dashed line to $\tau_p = 2$; and
the dot-dashed line to $\tau_p = 3$.

\noindent Figure 4abcd \qquad The principal physical quantities as functions of
$l_h$ for $l_s/l_h = 0.1$.  The different curves correspond to different
values of $\tau_p$, as described in the caption to Figure 3.

\noindent Figure 5abcd \qquad The principal physical quantities as functions of
$l_h$ for $l_s/l_h = 0.01$.  The different curves correspond to different
values of $\tau_p$, as described in the caption to Figure 3.

\noindent Figure 6abcd \qquad The relative contributions of the different
cooling mechanisms for $l_s/l_h = 1$.  The solid line is bremsstrahlung
(the sum of
e-e, e-p and e-e$^+$); the broken line is total inverse Compton (on both
internally and externally created photons); the dotted line is annihilation
cooling.  In some cases annihilation cooling is so weak that it accounts
for less than $10^{-5}$ of the total, and its curve does not appear.

a) $\tau_p = 0.1$

b) $\tau_p = 0.5$

c) $\tau_p = 1$

d) $\tau_p = 2$

\noindent Figure 7abcd \qquad The relative contributions of the different
cooling mechanisms for $l_s/l_h = 0.1$.  The mechanisms associated with
each curve are as described in the caption to Figures 6.

a) $\tau_p = 0.1$

b) $\tau_p = 0.5$

c) $\tau_p = 1$

d) $\tau_p = 3$

\noindent Figure 8abcd \qquad The relative contributions of the different
cooling mechanisms for $l_s/l_h = 0.01$.  The mechanisms associated with
each curve are as described in the caption to Figures 6.

a) $\tau_p = 0.1$

b) $\tau_p = 0.5$

c) $\tau_p = 1$

d) $\tau_p = 3$

\noindent Figure 9 \qquad Temperature in the Compton-dominated regime
as a function of $\tau_T$ for a variety of seed photon supply rates.
The symbol coding is: $l_s/l_h = 1$ represented by filled circles;
$l_s/l_h = 0.5$ by solid squares; $l_s/l_h = 0.1$ by crosses; $l_s/l_h = 0.01$
by open circles; and $l_s/l_h = 0.001$ by open triangles; the open square
represents an equilibrium corresponding to $l_s/l_h =10$.

\noindent Figure 10 \qquad The scaling of Comptonization model parameters
with $l_s/l_h$.
In these figures we have plotted results representative of all
the models we computed.

a) $\Theta \tau_T$ as a function of $l_s/l_h$.  Despite differences in
$l_h$ and $\tau_p$, all models fall very close to a straight line in
$\log (\Theta\tau_T)$ vs. $\log l_s/l_h$.  For fixed $\tau_p$, the spread
in $\tau_T$ is due to pair creation.

b) $\alpha$ as a function of $l_s/l_h$.  Although three independent parameters
are required to specify these models ($l_h$, $l_s$, and $\tau_p$), the
spectral index of the power-law component of the output spectrum depends
almost solely on one parameter, $l_s/l_h$.  When $l_s/l_h \simeq 1$, there
is a weak secondary dependence on $\tau_T$.

\end{document}